\documentclass[10pt,twocolumn,letterpaper]{article}
\usepackage{array}
\usepackage{wacv}
\usepackage{times}
\usepackage{epsfig}
\usepackage{graphicx}
\usepackage{amsmath}
\usepackage{amssymb}
\usepackage{booktabs}
\usepackage{subcaption}
\usepackage{multirow}
\wacvfinalcopy 

\ifwacvfinal
\usepackage[breaklinks=true,bookmarks=false]{hyperref}
\else
\usepackage[pagebackref=true,breaklinks=true,colorlinks,bookmarks=false]{hyperref}
\fi

\begin{document}

\title{End-to-end Rain Streak Removal with RAW Images}
\author{First Author GuoDong DU\\
Institution1:   National University of Singapore\\
Institution1 address:   21 Lower Kent Ridge Rd, 119077\\
{\tt\small duguodong7@gmail.com}
\and
Second Author HaoJian Deng\\
Institution2 Harbin Institute of Technology, Shenzhen\\
First line of institution2 address:  Harbin Institute of Technology, Shenzhen \\
\and
Third author JiaHao Su\\
Institution3: Harbin Institute of Technology, Shenzhen\\
First line of institution3 address: Harbin Institute of Technology, Shenzhen \\
\and
Fourth author Yuan Huang\\
Institution4: Waseda University\\
First line of institution4: address West Waseda 1-6-1, Shinjuku ku, Tokyo \\
\and
}

\maketitle
\thispagestyle{empty}

\begin{abstract}
   In this work we address the problem of rain streak removal with RAW images. The general approach is firstly processing RAW data into RGB images and removing rain streak with RGB images. 
Actually the original information of rain in RAW images is affected by image signal processing (ISP) pipelines including none-linear algorithms, unexpected noise, artifacts and so on. 
It gains more benefit to directly remove rain in RAW data before being processed into RGB format.
To solve this problem, we propose a joint solution for rain removal and RAW processing to obtain clean color images from rainy RAW image. 

[Method]
To be specific, we generate rainy RAW data by converting color rain streak into RAW space and design simple but efficient RAW processing algorithms to synthesize both rainy and clean color images. 
The rainy color images are used as reference to help color corrections. 

[Result]
Different backbones show that our method conduct a better result  compared with several other state-of-the-art deraining methods  focused on color image. In addition, the proposed network generalizes well to other cameras beyond our selected RAW dataset. Finally, we give the result tested on images processed by different ISP pipelines to show the generalization performance of our model is better compared with methods on color images. 
\end{abstract}

\section{Introduction}

\label{sec:intro}
As a fundamental problem in computer vision, rain removal aims to reconstruct the visibility of objects and scenes in photos captured under rainy days.  Since the atmospheric process is complicated, the visibility degradation caused by rain have different types in terms of raindrop size, rain density, and wind velocity~\cite{mukhopadhyay2014combating}. For example, nearby rain streak can distort background scene content; adherent raindrops that fall and flow on camera lenses can obstruct or blur the scenes; distant rain accumulated throughout the scene, creating a mist-like phenomenon in a manner more similarly to fog. Rain removal has thus become a necessary preprocessing step for subsequent outdoor tasks, like visual tracking ~\cite{song2018vital}, person reidentification~\cite{mao2017can}, or event detection~\cite{shehata2008video}. 

In the last decade, we have witnessed a continuous progress on rain removal research with many methods for both video and images. While conventional methods utilize the prior knowledge of background scenes and encoding physical properties, 
most recent methods employ the data-driven manner and design specific convolution neural network architectures. Some works focus on better visual quality using the generative adversarial network\cite{zhang2019image}, others focus on rain removal on the real world rain/rain-free image pairs besides on synthesized dataset\cite{spanet}.
All existing methods remove rain for images in RGB format which are processed from original RAW images. However there is significant information loss when RAW data is processed into RGB format through image signal processing (ISP) pipeline. 
No existing methods explore whether the original information in RAW image is helpful for rain removal tasks.  Not only the rainy image suffer from degradation of scene visibility outside the camera, but also it is affected by ISP pipeline within cameras. The main affection of ISP in summarized as below.
Firstly, the non-linear operation give fluctuated digit gains to the RAW pixels within a single image. For example, lens correction gives a bigger gain to the pixels away from the image center\cite{LensC}; 
Gamma correction  makes the dark regions lighter\cite{gammaC}; Tone mapping gives more contrast to the moderate pixels\cite{Reinhard_tonemp}.
These non-linear operations process the RAW data to match human vision but  are dishonest to the original rain streak distribution in linear RAW data space. 
In addition, camera's exposure algorithm usually gives the output color images a biased digit gain without considering image intensities. 
Secondly, the demosaicing process within ISP pipeline may produce construction noise, blurry and ringing artifacts\cite{Trinity}. 
These unexpected image quality degradation caused by digital image processing may also deteriorate the feature learning ability for deraining problems. 
Finally, the pixels in RAW data are typically 12 or 14 bits, but are reduced to 8 bits color pixels. The information loss brings more difficulties especially for high frequency learning tasks. 
Hence it is conspicuous that the ISP pipelines surely affect  rain removal tasks and no existing methods remove rain directly in RAW images. 
Furthermore, because different camera factories have different RAW processing pipelines, the final output RGB images exhibit different ISP characteristics. 
Thus the learned deraining model from a kind of ISP images may generalize poorly on color images processed by another factory. A deep learning model always needs a new dataset for different camera factories. 
While the domain distance of different RAW images is shorter than color images, the rain removal architectures proposed for color images are not general enough to model the difference between ISP pipelines.  

The RAW data from cameras is accessible easily, and significant progress has been made for image restoration and enhancement by exploiting the RAW data in some low-level vision tasks.
These tasks are devoted to solve the degradation reconstruction problems such as denoising~\cite{RAWDN}, super resolution~\cite{RAWSR} and illumination enhancement~\cite{SID}. 
In this work, we attempt to exploit RAW data information to further boost rain removal performance based on deep network architectures which are prominent for strong representation capability. Specially, we propose a joint solution for rain removal and RAW processing by designing a network architecture. What is more, to obtain paired RAW rainy images and clean RGB images, we design algorithms to model rain streak in RAW domain and synthesize RAW rainy images. 

\begin{figure}[htbp]
  \centering
  \setlength{\fboxsep}{0pt} 
  \setlength{\fboxrule}{1pt} 
  \fbox{\includegraphics[width=8cm, height=4.45cm]{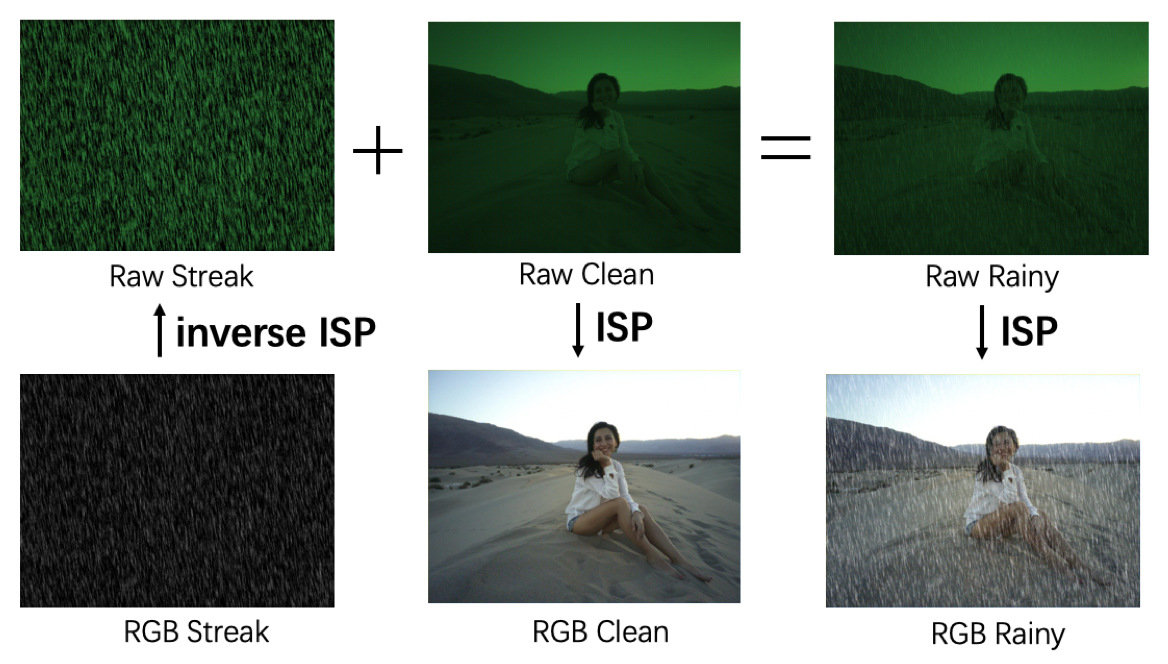}}
  \caption{RAW rainy image pipeline.}
  \label{RAW rainy image pipeline.}
\end{figure}

\begin{figure*}[htbp]
  \centering
  \setlength{\fboxsep}{0pt} 
  \setlength{\fboxrule}{1pt} 
  \fbox{\includegraphics[width=16cm, height=8cm]{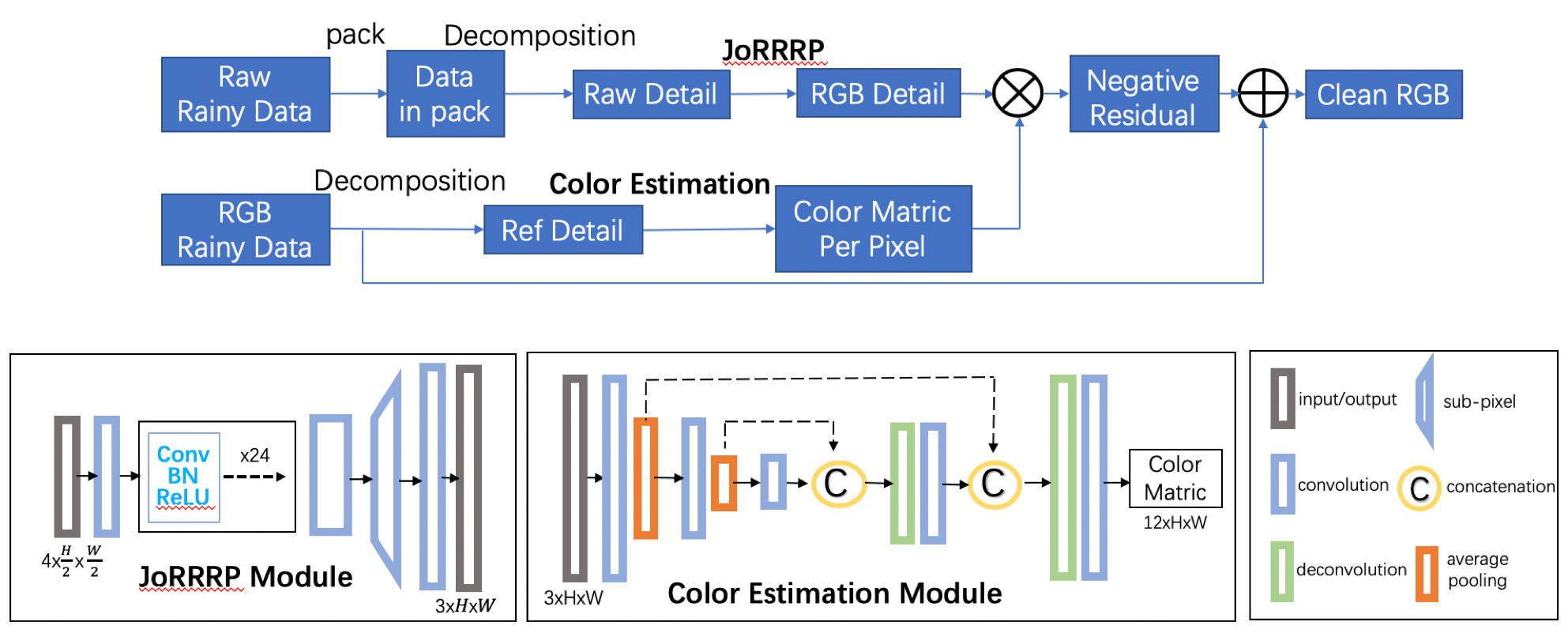}}
\caption{Joint Rain Removal and RAW Processing.}
\label{Joint Rain Removal and RAW Processing.}
\end{figure*}

To synthesize rainy images in RAW data and reserve realistic details, we firstly select a high quality RAW image dataset as our clean or background RAW images. Since we mainly focus on the rain removal task in RAW images, we model the rain image considering only the type of rain streak. We render RAW rain streaks which are then added to the clean RAW images to synthesize a rainy RAW image. RAW rain streaks are modeled by converting rain streaks in color space, using the inverse ISP technology, which is proposed by \cite{RAWDN}. Then we design a simple but efficient ISP pipeline processing both the RAW clean and rainy data to obtain corresponding images in RGB format separately. Our pipeline for rainy RAW image synthesis is shown in figure  \ref{RAW rainy image pipeline.}.
Besides, we choose another two RAW image datasets and implement several different ISP pipelines to establish a benchmark for generalization ability test.  
For the\textbf{ Jo}int \textbf{R}ain \textbf{R}emoval and \textbf{R}aw \textbf{P}rocessing network (JoRRRP) architecture, we first pack RAW rainy data according to its bayer pattern and then extract a RAW detail layer obtained by using low-pass filtering\cite{Bilateral,DDN}.  
The layer is then fed into deep residual network (ResNet) \cite{ResN} followed by an upsampling layer using pixel shuffle method\cite{PS}. Hence we can obtain a demosaiced color rain mask. Then we apply a spatially-variant color transformation estimated from color rainy image to help recover the fidelity of color appearances.  Thus we obtain the rain mask in RGB format which is the negative residual mapping as proposed in \cite{DDN}. After combining the rain mask in RGB format with processed rainy RGB image, we can get the final clean RGB image. Our overall framework for end-to-end rain streak removal in RAW images is shown is Figure \ref {Joint Rain Removal and RAW Processing.}.

We conduct experiments on our RAW synthetic rain data to demonstrate the importance of each component of our model. In addition, we replace ResNet with other efficient backbones\cite{RDN, RCAN, IMDN} for further ablation study. The result shows that the performance of rain streak removal with RAW images is always better than RGB methods, with average 0.7 dB and 0.05 improvement, according to the evaluation metric of peak signal-to-noise ratio (PSNR) and structural similarity (SSIM) index separately. Finally, we give the result tested on images processed by different ISP pipelines to show the generalization performance of our model compared with methods on color images. 
Our contributions are summarized as follows:
\begin{enumerate}
\item We present a technique to synthesize rainy image on RAW data, especially a pipeline of how to generate rain streak on RAW data. To the best of our knowledge, this is the first dataset that is aimed to use RAW data for deraining problems.
\item We propose a joint rain removal and RAW processing deep learning model. Experiments show that using RAW data achieves better accuracy and visual improvements against state-of-the-art methods on only RGB space images.
\item Extensive experiments are conducted on images from different ISP pipelines and results show that our me-thod gives a better generalization performance.

\end{enumerate}
\section{Related Work}
Our work is related to RAW image processing and rain removal methods, which are reviewed briefly as follows.


\textit{A. RAW image processing} 

Many joint solutions for image restoration tasks using deep neural network have been proposed by exploiting RAW data. \cite{dmdn} proposed an efficient CNN for effective joint denoising and demosaicking.
\cite{Trinity} address demosaicing and super-resolution simultaneously and claim that super-resolution with RAW data helps recover fine details and clear structure.
\cite{RAWDN} present procedures for unprocessing color images into synthetic RAW data and the purpose is to obtain more realistic noisy and clean pairs. The problem of lack of realistic training data also exists in super resolution. \cite{RAWSR} develop a dual convolutional neural network to exploit the originally captured radiance information in RAW images and they simulate the imaging process of digital cameras to remedy the information loss of the input.  
Their proposed method is declared to enable super-resolution for real captured images. \cite{SID} introduce a dataset of RAW short-exposure low-light images, with corresponding long-exposure reference images and  develop a network for processing low-light images.  \\
There exists a few datasets with different imaging scenarios that can provide metadata of camera sensors for rendering RAW rain streak. can be used as clean background RAW data. The HDR+ dataset is featured with burst denoising and sophisticated style retouching \cite{Reinhard_tonemp}; 
\cite{Fairchild} proposed an unique database of HDR photographs accompanied by detailed luminance measurements and visual appearance scaling from the original scene; 
and the FiveK dataset contains high quality images that are retouched by five photographers to have different color styles\cite{fivek}.   \\\\
\textit{B. Rain removal methods} 

When a rainy image is taken, the visual effects of rain on that digital image further hinges on many camera parameters, such as exposure time, depth of field, and resolution\cite{Garg}.  
Garg and Nayar proposed a camera based rain gauge and improve the performance of deraining by selecting camera parameters without appreciably altering the scene appearance. 
Besides, the author claimed that the visibility of rain is affected by both the brightness of scene and camera parameters such as exposure time and depth of field.\\
While multi-frame based deraining method \cite{garg2004detection, kim2015video} take the advantages of sequential information, the prior based algorithms for single image deraining are more available.  Chen and Hsu \cite{chen2013generalized} decomposed the background and rain streak layers based on low-rank priors. 
Li et al. \cite{li2016rain} use patch-based priors for both the clean background and rain layers in the form of Gaussian mixture models. These approaches tend to have unsatisfactory performances on real images with complicated scenes and rain forms. 
Numerous deep learning techniques for image deraining algorithms have been recently proposed. 
Yang et al. attempt to jointly detect and remove rains taking a contextualized dilated network\cite{jorder}. 
Zhang et al. apply a residual-aware classifier to efficiently determine the density-level of a given rainy image for joint rain density estimation and deraining\cite{didmdn}. Wang et al. design a novel SPANet in a local-to-global attentive manner\cite{spanet}. 
None of the existing methods exploit the RAW data for rain removal tasks.\\

\section{Method}
\subsection{RAW Image Pipeline}
Most existing deraining datasets are synthesized by assuming a specific rain model, such as rain streak, raindrop, as well as rain and mist.  The most common model assumed by the majority of deraining algorithms is a rain streak image $R_{s}$, 
which is modeled as a linear super- imposition of the clean background scene $B$ and the sparse, line-shape rain streak component $S$:  $R_{s} = B + S$
Since our research mainly concern how the RAW data information could help for deraining problems, we only use this certain type of synthetic images. Then we explore the technique to "unprocess" rain streak to its corresponding RAW format 
by inverting each component in RAW processing pipeline. The original RAW data will be used as background of our rainy RAW image, and we generate rain streak according to the clean RAW data.\\

Datasets such as MiT-Adobe Fivek \cite{fivek} record the camera metadata of their images which is very import when converting RAW data to RGB space by image signal processing (ISP). 
The metadata usually comprises important information such as black level, saturation , Bayer pattern, color matrix, flip and so on. Given the metadata of each RAW data image,
we can process our rain streak to its corresponding RAW rain streak. See Figure  \ref{RAW rainy image pipeline.} for an overview of our steps of rain streak generation on RAW data. \\

A traditional ISP pipeline usually contains denoising, lens correction, exposure white balance, color correction, tone mapping, gamma correction and so on.  
We design a simple but efficient ISP pipeline to process RAW image, then we inverse each step of this pipeline for rendering RAW rain streak.

\begin{figure*}[htb]
  \centering
  \begin{subfigure}[t]{0.32\linewidth}
    \centering
    \includegraphics[width=0.99\linewidth]{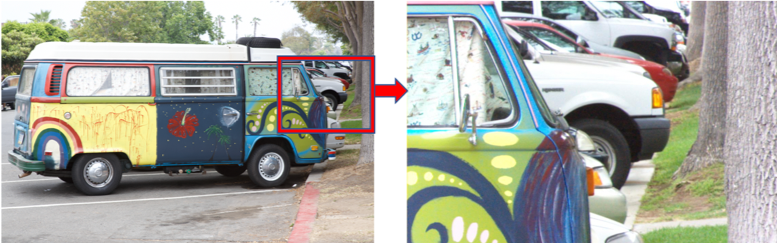}
    \caption*{RGB clean.}
  \end{subfigure}
  \begin{subfigure}[t]{0.64\linewidth}
    \centering
    \includegraphics[width=0.23\linewidth]{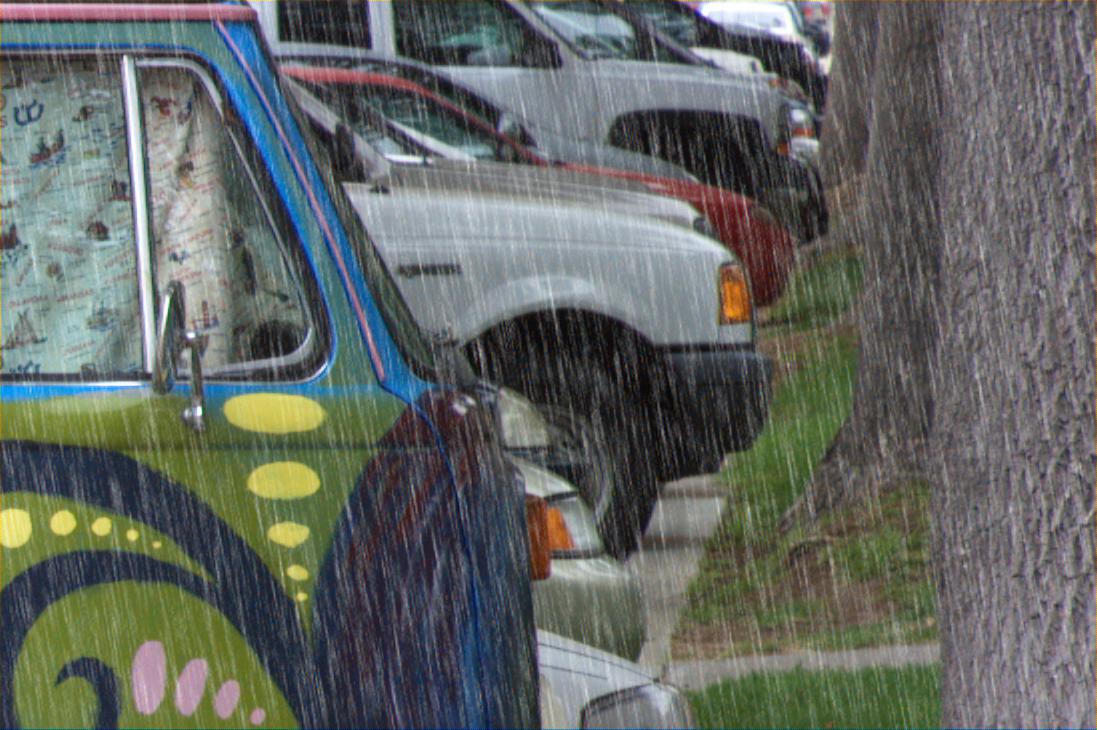}
    \includegraphics[width=0.23\linewidth]{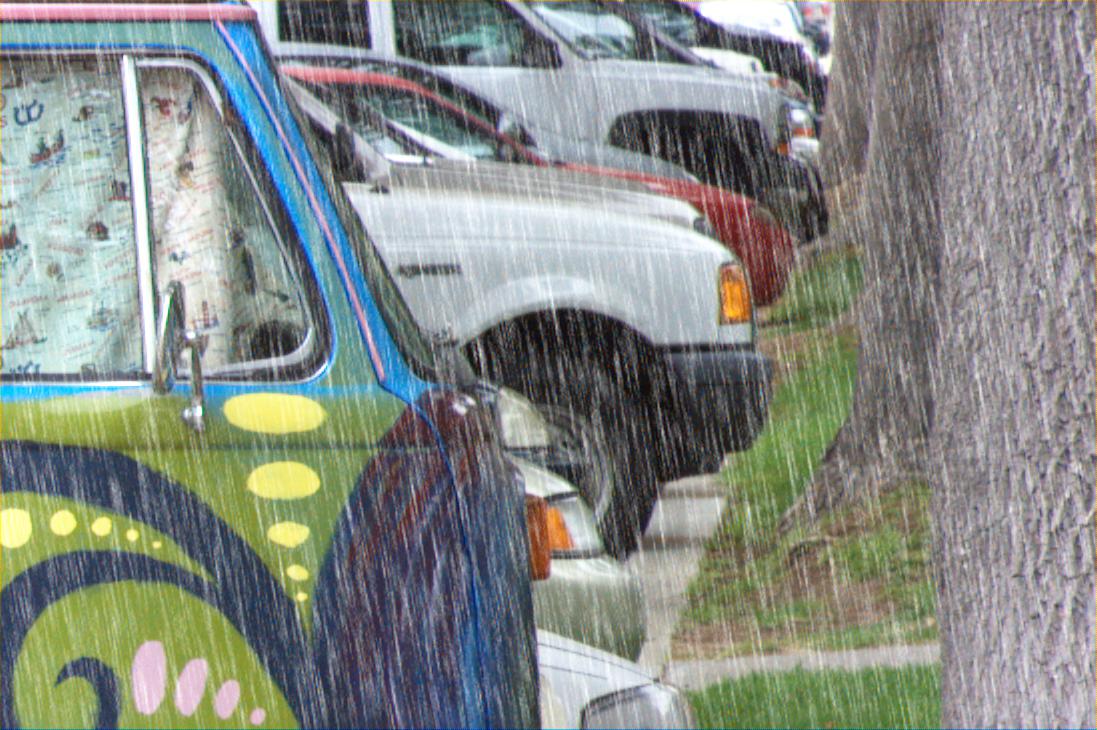}
    \includegraphics[width=0.23\linewidth]{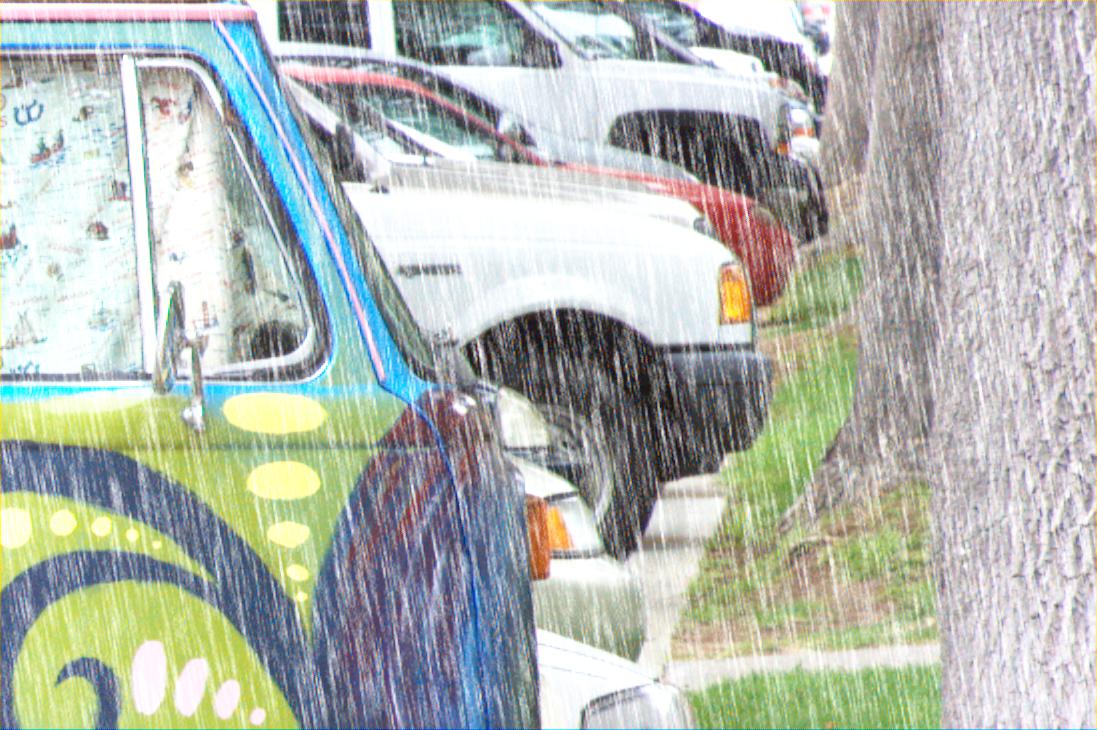}
    \includegraphics[width=0.23\linewidth]{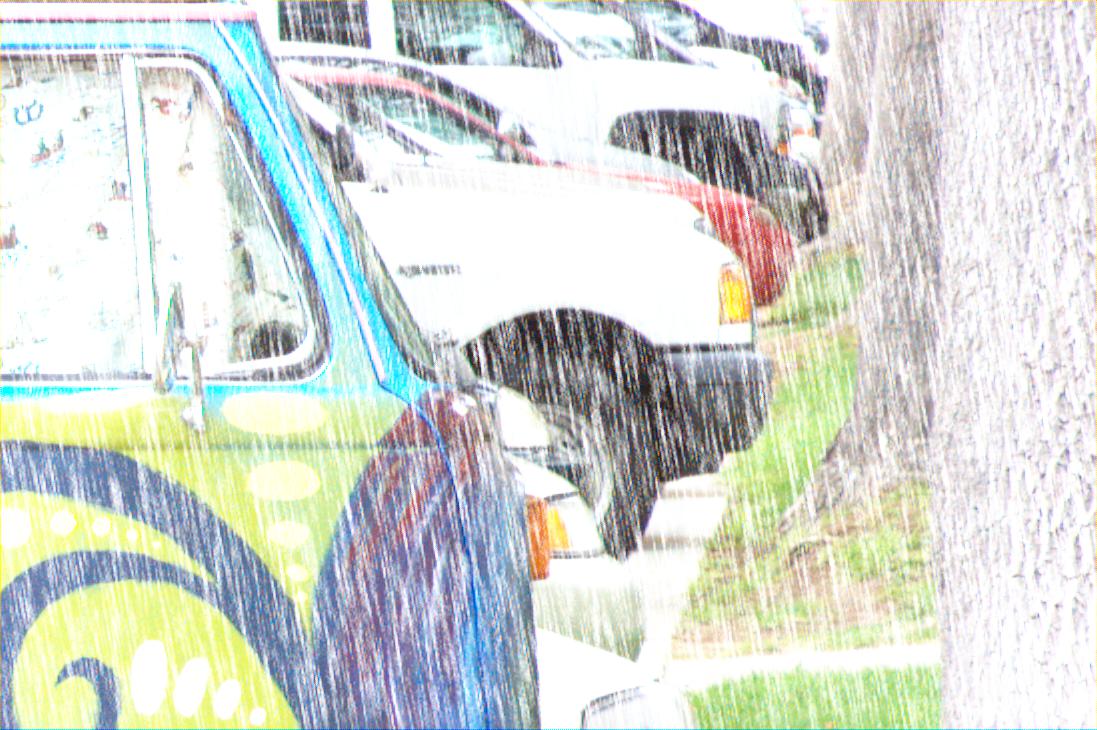}
    \caption*{exposure}
  \end{subfigure}
  \begin{subfigure}[t]{0.32\linewidth}
    \centering
    \includegraphics[width=0.99\linewidth]{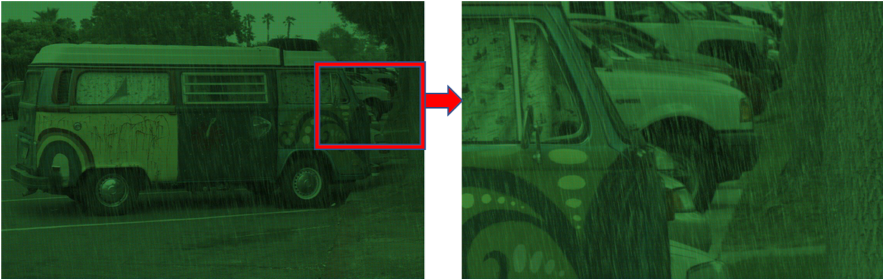}
    \caption*{RAW rainy}
  \end{subfigure}
  \begin{subfigure}[t]{0.15\linewidth}
    \centering
    \includegraphics[width=0.99\linewidth]{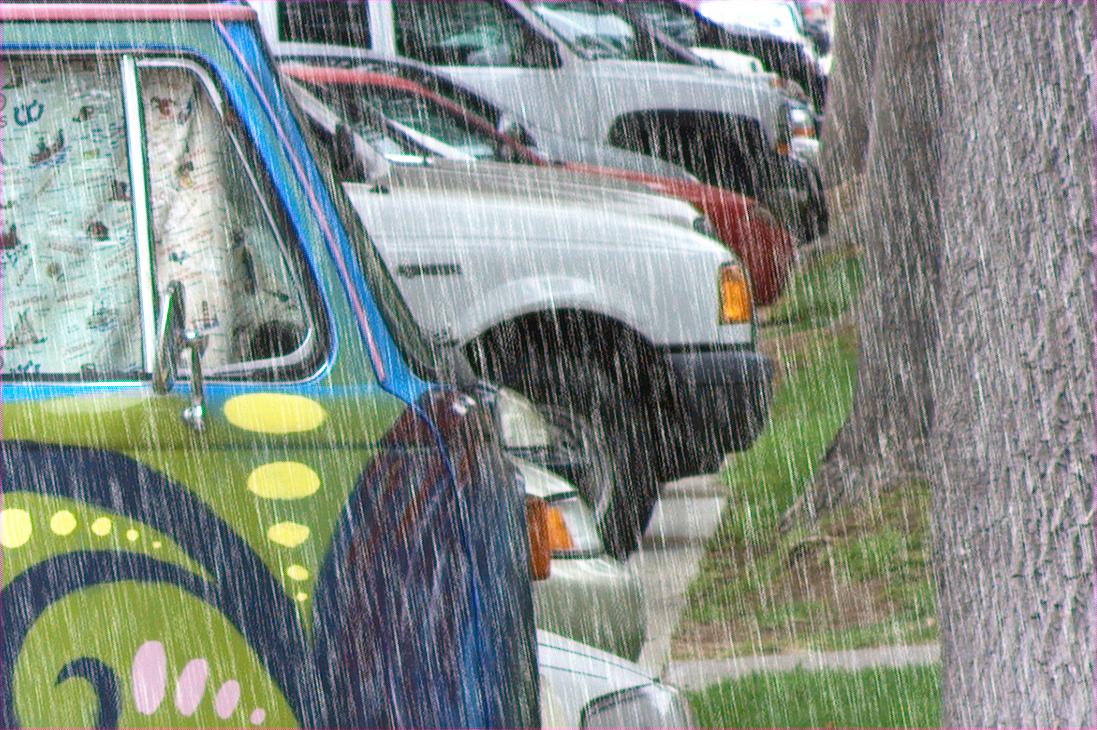}
    \caption*{demosaic}
  \end{subfigure}
  \begin{subfigure}[t]{0.46\linewidth}
    \centering
    \includegraphics[width=0.32\linewidth]{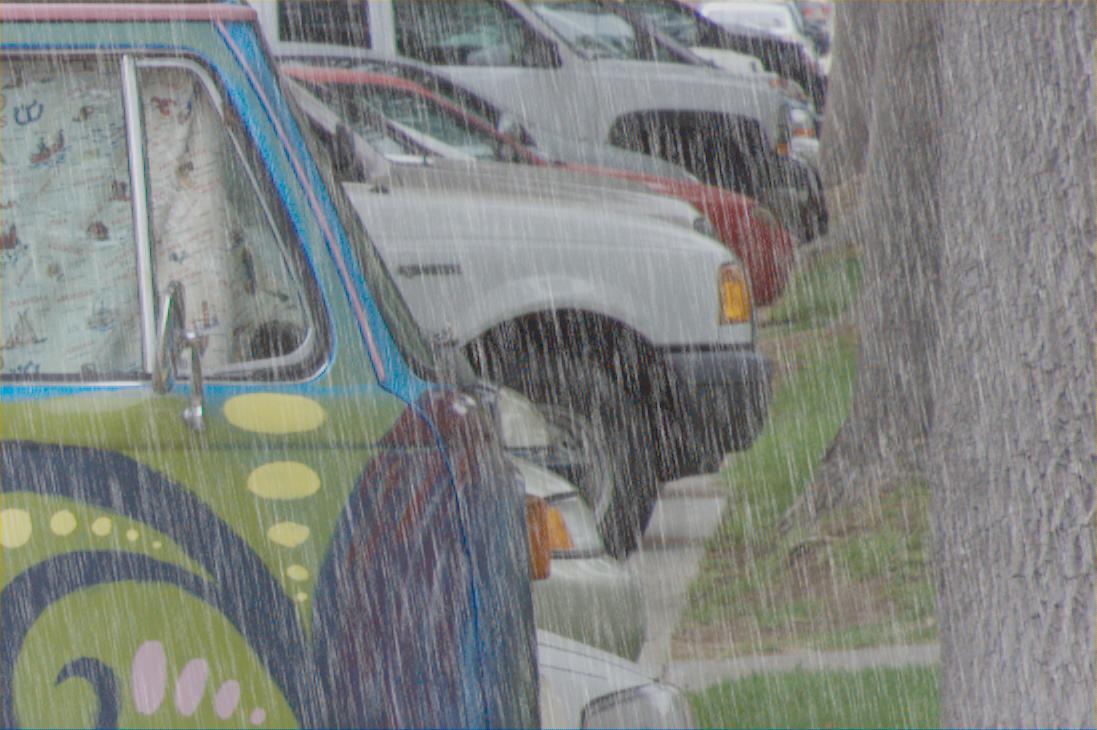}
    \includegraphics[width=0.32\linewidth]{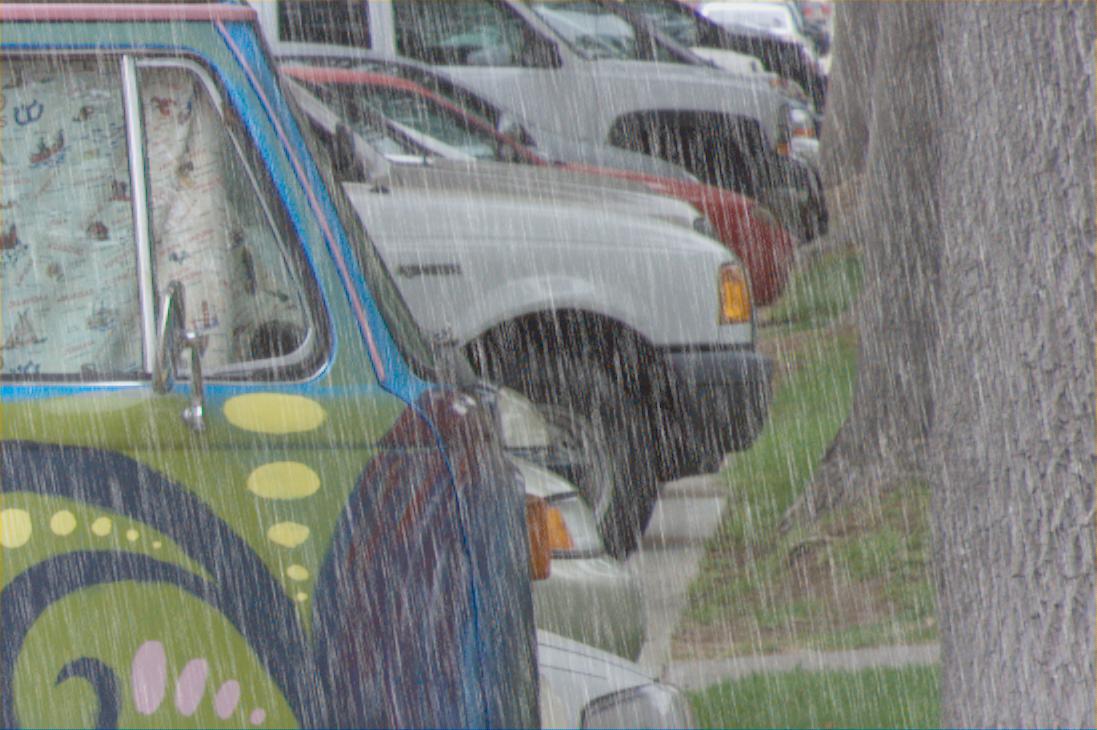}
    \includegraphics[width=0.32\linewidth]{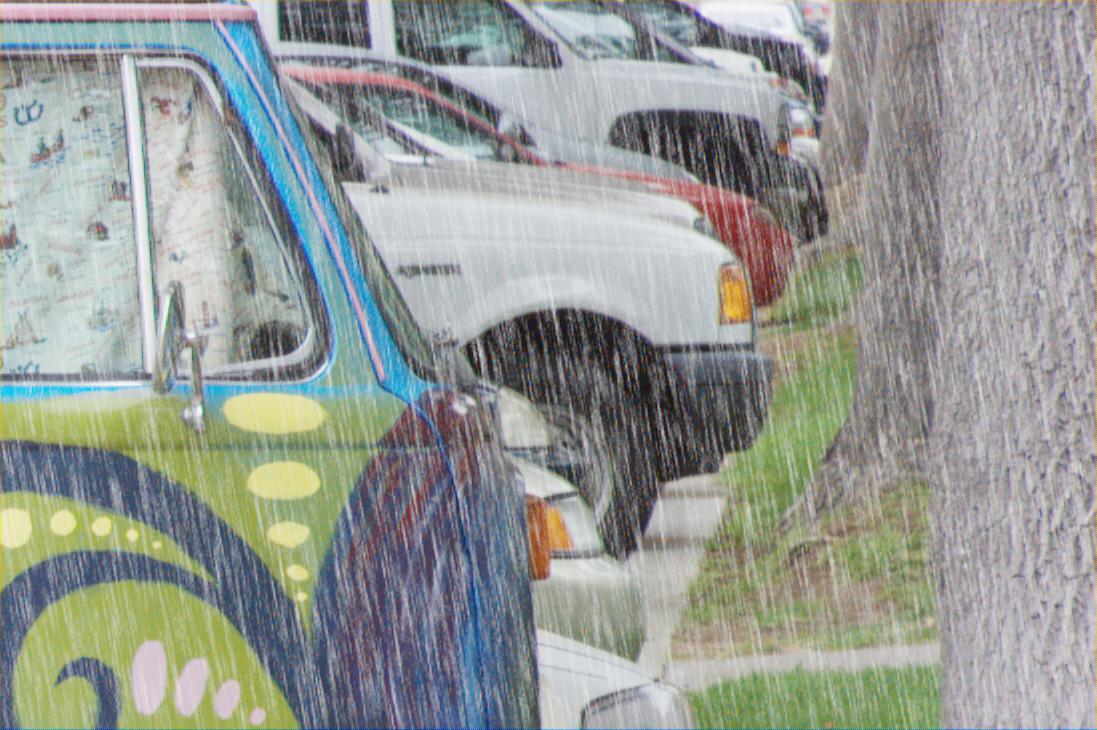}
    \caption*{tone mapping}
  \end{subfigure}

  \caption{Different ISP pipelines}
  \label{Different ISP pipelines}
\end{figure*}

What is more, we propose several different RAW processing algorithms simulating different styles of ISP to obtain various color images for testing of generalization performance.  See Figure \ref {Different ISP pipelines} for comparison of different ISP processing results.\\\\
\textbf{Noise and Lens Correction:} 

Our selected RAW sensor data actually have its real noise, primarily coming from two sources: photo arrival statistics (“shot” noise) and imprecision in the readout circuitry (“read” noise). 
Hence we just assume the generated rainy RAW data keep the same noise as its original clean background, it is trivial to consider how to imitate RAW noise in rain streak.  
The lens profile correction adjustments is to remove typical optical issues that camera lenses create, such as vignetting or distortion\cite{LensC}. 
We assume a simple radial distortion lens correction function: 
$$ RAW_{corrected} = RAW \cdot \left(\alpha r^{2} + \beta \right ) $$
where $r$ is the distance from image center to each pixel. $\alpha, \beta$ are fluctuated constant for different color channels.
For RAW rain streak generation, we inverse this step through division by gain map. \\\\
\textbf{Demosaicing:}

In the beginning of digital image processing pipeline we need to reconstruct a full color image from the incomplete color samples output from an image sensor overlaid with a color filter array (CFA), such as R-G-G-B.
This can be achieved by interpolation or more complicated demosaicing algorithms [15]. To invert this step we omit two of its three color values according to the Bayer filter pattern for each pixel in the our rain streak images.
In our forward image signal processing (ISP) of rendering RGB images from both clean and rainy RAW data, we adopt a demosaic algorithm with linear interpolation for each color pixel to perform color construction. 
For test dataset, we adopt another advance demosaicking method using finite impulse response filter in frequency domain to extract necessary components.\\\\
\textbf{Exposure:}

As a extremely simple brightening measure, we find the mean luminance of the image and then scale it so that the mean luminance is some more reasonable value. 
We fairly arbitrarily scale the image so that the mean luminance is about 1/4 the maximum, with a normal distribution centered at 0.25 with standard deviation of 0.05, and then clamped into range $\left[0.045, 0.75\right]$ for a safe content reserving. For the ISP pipelines designed for test benchmark, we choose a larger range of mean luminance values.\\\\
\textbf{White Balance:}

It is very important to estimate illumination chromaticity correctly to avoid invalid overall color cast in the final image.
The goal of white balancing is to estimate accurately the color of the overall scene illumination and to make the image look as if is taken under canonical light. When synthesizing rain streak on RAW data, 
we take the product of the inverse digital and white balance gains to get an inverse gain for per-channel, then we  apply this gain to our rain streak. This inverse gain is almost always less than unity, 
which means that naively gaining down our synthetic imagery will result in a dataset that systematically lacks highlights and contains almost no clipped pixels. 
This is problematic, as correctly handling saturated image intensities is critical when generating rain streak on RAW data. This problem is the same in \cite{RAWDN} 
when the author investigates the pipeline of generation of realistic synthetic RAW data for denoising.  The author then apply a highlight-preserving transformation $ f(x, g) $ to perform safe white balance inverse. 
\begin{align}
a(x) &= \left ( \frac{\max\left ( x-t, 0 \right )}{1-t} \right )^{2}  \\
f(x,g) &= \max\left ( \frac{x}{g}, \left ( 1-\alpha \left ( x \right ) \right )\left ( \frac{x}{g} \right )+\alpha \left ( x \right )x\right)  
\label{equation2}
\end{align}
We apply this function for our rain streak generation problem. \\\\
\textbf{Color Correction:}

In general, the color space in a camera space is different from expected sRGB color space because the color filters of a camera sensor vary from different factories. 
Then the camera RAW data applies a 3 x 3  color correction matrix (CCM) to address this problem. Thus we apply the inverse of a RAW data CCM to transmit a rain streak from sRGB space to the original “camera space”. \\\\
\textbf{Gamma Compression:}

Gamma Compression is is a nonlinear operation used to make dark regions lighter by allocating more bits of dynamic range to low intensity pixels, 
because humans are more sensitive to relative differences between darker tones than between lighter tones. 
\begin{align}
\Gamma \left ( x \right ) &= \max(x, \epsilon )^{\frac{1}{2.4}}     \\
\Gamma^{-1}\left ( y \right ) &= \max(y, \epsilon )^{2.4}      
\end{align}\\
\textbf{Tone Mapping:}

Tone mapping addresses the problem of strong contrast reduction from the scene radiance to the displayable range while preserving the image details and color appearance important to appreciate the original scene content. 
It is conspicuous that RAW data are always high dynamic range images and processing it to color images prefer a well designed tone mapping algorithm\cite{Reinhard_tonemp}. 
The reverse-engineering for a complex edge-aware local tone mapping is difficult and trivial. We choose a simple tone mapping curve the same as equation \ref{equation5} and inverse the curve function for rain streak unprocessing.
\begin{align}
\label{equation5}
simple(x) &= 3x^{2}-2x^{3}             \\
simple^{-1}(y) &= \frac{1}{2} - \sin \left ( \frac{\sin^{-1}\left ( 1-2y \right )}{3} \right )   
\end{align}
both are only defined on inputs in [0, 1]. \\
\begin{minipage}{\textwidth}
\begin{minipage}{0.49\linewidth}
\centering
\includegraphics[width=0.98\textwidth]{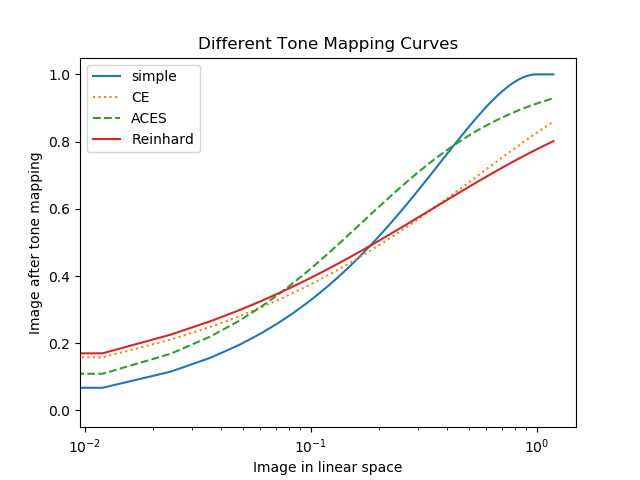}
\makeatletter\def\@captype{figure}\makeatother
\caption{tone mapping}
\label{tone mappping}
\end{minipage}
\vspace{0.3cm}
\hfill\\
\begin{minipage}{0.48\linewidth}
\centering
\begin{align}
Reinhard\left ( x \right )   &= \frac{x}{1+x}      \\
CE\left ( x \right )   &= 1 - e^{-x} \\
ACES\left ( x \right )   &=  \frac{x (2.51x + 0.03))}{x (2.43x + 0.59) + 0.14}
\end{align}
\vspace{0.4cm}
\end{minipage}
\end{minipage}

For the ISP pipelines designed for test benchmark, we choose another 3 common tone mapping method: Reinhard tone mapping, Cry Engine 2 tone mapping(CE), 
 the Academy Color Encoding System (ACES) tone mapping, \cite{Reinhard_tonemp,CE_tonemp,ACES_tonemp}. 
 See Figure \ref{tone mappping} for the comparison of different tone mapping curves.\\
 
\subsection{Joint Rain Removal and RAW Processing}
Figure \ref{tone mappping} illustrates the proposed architecture and comprises two modules: Joint Rain Removal and RAW Processing (JoRRRP) module to learn a mapping function from RAW details to color details; Color estimation for predicting a local transformation matrix. 
\begin{figure*}[htbp]
\centering
\includegraphics[ width=14cm, height=7cm,angle=0, scale=1]{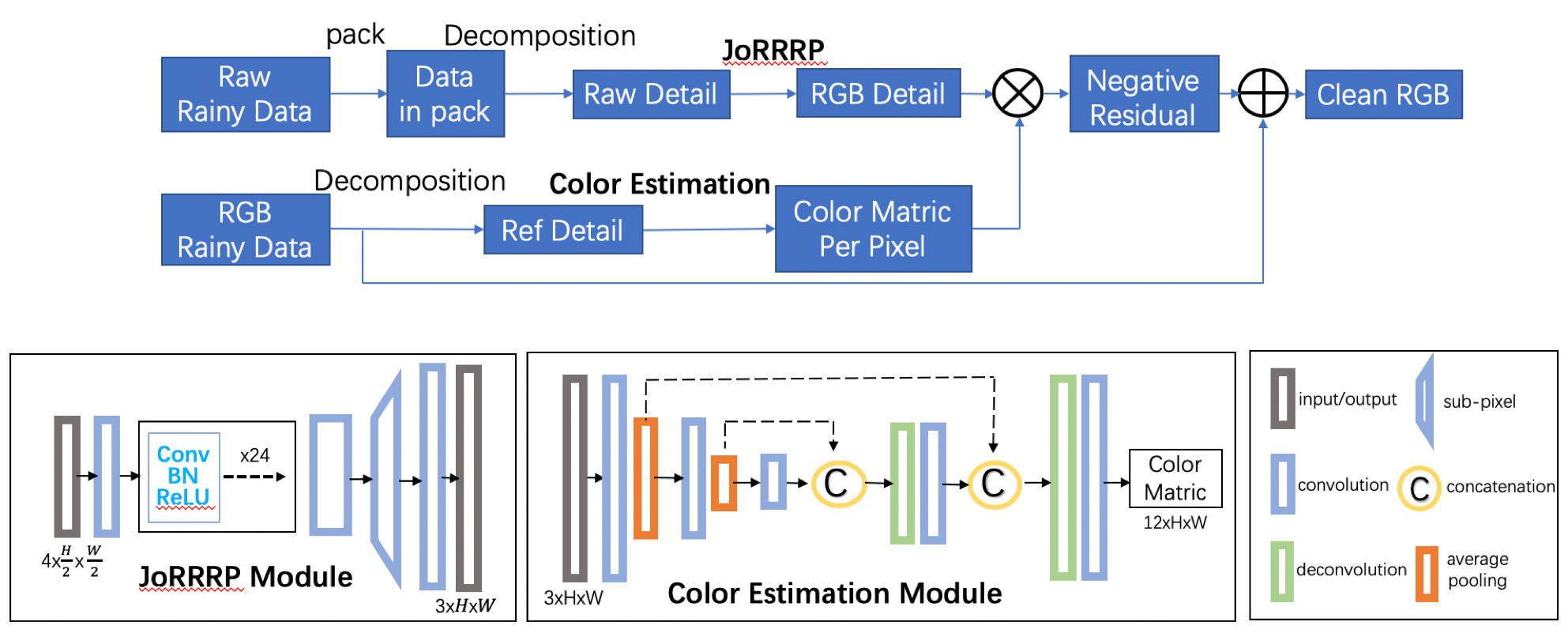}
\caption{Jodd}
\label{Jodd}
\end{figure*}
\begin{table*}[!htbp]
  \setlength{\belowcaptionskip}{0.2cm}
  \centering
  \caption{Ablation study for our joint RAW processing and deraining network}
\begin{tabular}{>{\centering\arraybackslash}m{3.8cm}|>{\centering\arraybackslash}m{1.0cm}>{\centering\arraybackslash}m{1.1cm}|>{\centering\arraybackslash}m{1.0cm}>{\centering\arraybackslash}m{1.1cm}|>{\centering\arraybackslash}m{1.0cm}>{\centering\arraybackslash}m{1.1cm}}
    \hline
    \multirow{2}{*}{Methods}& \multicolumn{2}{c|}{MIT-Adobe} & \multicolumn{2}{c|}{HDR+} & \multicolumn{2}{c}{Fairchild}    \tabularnewline 
                               &  \multicolumn{1}{c}{PSNR}  & \multicolumn{1}{c}{SSIM} & \multicolumn{1}{|c}{PSNR} & \multicolumn{1}{c}{SSIM} & \multicolumn{1}{|c}{PSNR} & \multicolumn{1}{c}{SSIM} \tabularnewline
    \hline
	  only rgb input  & 28.95  & 0.9174  & 28.41  & 0.9320  & 28.60  & 0.9373 \tabularnewline
w/o color branch   & 27.28  & 0.8858  & 26.42  & 0.8798  & 26.96  & 0.9056 \tabularnewline
w/o RAW input, w/o pack  & 28.73  & 0.9182  & 28.67  & 0.9333  & 28.70  & 0.9366 \tabularnewline
w/o RAW input, pack+DC  & 29.07  & 0.9225  & 29.03  & 0.9377  & 29.08  & 0.9413 \tabularnewline
w/o RAW input, DC+pack  & 29.04  & 0.9221  & 28.96  & 0.9374  & 28.94  & 0.9408 \tabularnewline
w/o feature fusion  & 29.51  & 0.9272  & 29.28  & 0.9422  & 29.20  & 0.9441 \tabularnewline
Ours full model  & \textcolor[rgb]{ 1,  0,  0}{29.53}  & \textcolor[rgb]{ 1,  0,  0}{0.9278}  &\textcolor[rgb]{ 1,  0,  0}{ 29.33}  &\textcolor[rgb]{ 1,  0,  0}{ 0.9431}  & \textcolor[rgb]{ 1,  0,  0}{29.25}  & \textcolor[rgb]{ 1,  0,  0}{0.9482}                    
    \end{tabular}%
\end{table*}%
\begin{table*}[!htbp]
  \setlength{\belowcaptionskip}{0.2cm}
  \centering
  \caption{comparison of backbones for joint rain removal and RAW processing.}
\newcolumntype{C}[1]{>{\centering\arraybackslash}p{#1}}
    \begin{tabular}{C{1cm}|C{2.4cm}|C{1.0cm}C{1.1cm}|C{1.0cm}C{1.1cm}|C{1.0cm}C{1.1cm}}
    \hline
    \multicolumn{2}{c|}{\multirow{2}{*}{Backbones}} & \multicolumn{2}{c|}{MIT-Adobe} & \multicolumn{2}{c|}{HDR+} & \multicolumn{2}{c}{Fairchild} \tabularnewline
    \multicolumn{2}{c|}{} & PSNR  & \multicolumn{1}{c|}{SSIM} & \multicolumn{1}{c}{PSNR} & \multicolumn{1}{c|}{SSIM} & \multicolumn{1}{c}{PSNR} & \multicolumn{1}{c}{SSIM} \tabularnewline 
    \hline
 \multirow{3}[2]{*}{ResN} & only rgb input & 28.95 & 0.9174 & 28.41 & 0.9322 & 28.59 & 0.9373  \tabularnewline
 & w/o RAW input & 29.07 & 0.9225 & 29.02 & 0.9377 & 29.07 & 0.9413  \tabularnewline
 & Ours full model & \textcolor[rgb]{ 1,  0,  0}{29.52} & \textcolor[rgb]{ 1,  0,  0}{0.9278} & \textcolor[rgb]{ 1,  0,  0}{29.32} & \textcolor[rgb]{ 1,  0,  0}{0.9431} & \textcolor[rgb]{ 1,  0,  0}{29.25} & \textcolor[rgb]{ 1,  0,  0}{0.9482}  \tabularnewline\hline
\multirow{3}[2]{*}{RDN} & only rgb input & 29.00 & 0.9229 & 28.50 & 0.9355 & 28.67 & 0.9407  \tabularnewline
 & w/o RAW input & 29.36 & 0.9261 & 29.00 & 0.9396 & 28.98 & 0.9424  \tabularnewline
 & Ours full model & \textcolor[rgb]{ 1,  0,  0}{29.94} & \textcolor[rgb]{ 1,  0,  0}{0.9332} & \textcolor[rgb]{ 1,  0,  0}{29.41} & \textcolor[rgb]{ 1,  0,  0}{0.9474} & \textcolor[rgb]{ 1,  0,  0}{29.49} & \textcolor[rgb]{ 1,  0,  0}{0.9520}  \tabularnewline\hline
\multirow{3}[2]{*}{RCAN} & only rgb input & 29.34 & 0.9263 & 29.24 & 0.9422 & 29.28 & 0.9466  \tabularnewline
 & w/o RAW input & 29.46 & 0.9270 & 29.34 & 0.9433 & 29.21 & 0.9460  \tabularnewline
 & Ours full model & \textcolor[rgb]{ 1,  0,  0}{30.80} & \textcolor[rgb]{ 1,  0,  0}{0.9339} & \textcolor[rgb]{ 1,  0,  0}{29.62} & \textcolor[rgb]{ 1,  0,  0}{0.9492} & \textcolor[rgb]{ 1,  0,  0}{29.54} & \textcolor[rgb]{ 1,  0,  0}{0.9532}  \tabularnewline\hline
\multirow{3}[2]{*}{IMDN} & only rgb input & 28.48 & 0.9205 & 28.91 & 0.9377 & 28.79 & 0.9390  \tabularnewline
 & w/o RAW input & 29.16 & 0.9239 & 29.03 & 0.9402 & 28.91 & 0.9420  \tabularnewline
 & Ours full model & \textcolor[rgb]{ 1,  0,  0}{29.74} & \textcolor[rgb]{ 1,  0,  0}{0.9303} & \textcolor[rgb]{ 1,  0,  0}{29.48} & \textcolor[rgb]{ 1,  0,  0}{0.9464} & \textcolor[rgb]{ 1,  0,  0}{29.48} & \textcolor[rgb]{ 1,  0,  0}{0.9508}  \tabularnewline\hline
    \end{tabular}%
\end{table*}%
\begin{table*}[!htbp]
  \setlength{\abovecaptionskip}{0.cm}
  \setlength{\belowcaptionskip}{0.2cm}
  \centering
  \caption{Experiments on different ISP pipelines.}
\newcolumntype{C}[1]{>{\centering\arraybackslash}p{#1}}
\begin{tabular}{C{1cm}|C{2.4cm}|C{1.0cm}C{1.1cm}|C{1.0cm}C{1.1cm}|C{1.0cm}C{1.1cm}}
    \hline
    \multicolumn{2}{c|}{\multirow{2}{*}{ISP pipelines}} & \multicolumn{2}{c|}{MIT-Adobe} & \multicolumn{2}{c|}{HDR+} & \multicolumn{2}{c}{Fairchild} \\
    \multicolumn{2}{c|}{} & RGB  & \multicolumn{1}{c|}{RAW} & \multicolumn{1}{c}{RGB} & \multicolumn{1}{c|}{RAW} & \multicolumn{1}{c}{RGB} & \multicolumn{1}{c}{RAW} \\ 
    \hline
     \multirow{3}[2]{*}{ResN} & simple   & 28.602 & +0.511 & 28.415 & +0.787 & 28.595 & +0.607\tabularnewline
& demosaic & 26.243 & +0.642 & 25.939 & +0.912 & 26.224 & +0.819\tabularnewline
& exposure & 27.864 & +0.635 & 27.511 & +0.993 & 27.854 & +0.944\tabularnewline
& tone mapping & 30.221 & +0.793 & 29.920 & +1.341 & 30.533 & +1.171\tabularnewline
\hline
 \multirow{3}[2]{*}{RDN} & simple   & 28.784 & +0.542 & 28.500 & +0.919 & 28.675 & +0.819\tabularnewline
& demosaic & 26.081 & +0.708 & 25.831 & +1.194 & 26.140 & +1.075\tabularnewline 
& exposure & 28.042 & +0.718 & 27.593 & +1.229 & 27.956 & +1.203\tabularnewline 
& tone mapping & 30.374 & +0.923 & 30.008 & +1.546 & 30.597 & +1.416\tabularnewline
    \end{tabular}%
\end{table*}%

\begin{table*}[!htbp]
   \setlength{\abovecaptionskip}{0.cm}
   \setlength{\belowcaptionskip}{0.2cm}
   \centering
   \caption{comparison with other rain removal methods.}
   \newcolumntype{C}[1]{>{\centering\arraybackslash}p{#1}}
\begin{tabular}{C{3.2cm}|C{1.0cm}C{1.1cm}|C{1.0cm}C{1.1cm}|C{1.0cm}C{1.1cm}}
     \hline
     \multirow{2}{*}{Methods}& \multicolumn{2}{c|}{MIT-Adobe} & \multicolumn{2}{c|}{HDR+} & \multicolumn{2}{c}{Fairchild}    \\ 
                                &  PSNR  & \multicolumn{1}{c|}{SSIM} & \multicolumn{1}{c}{PSNR} & \multicolumn{1}{c|}{SSIM} & \multicolumn{1}{c}{PSNR} & \multicolumn{1}{c}{SSIM} \\ 

     \hline
     DDN & 28.95 & 0.9174 & 28.42 & 0.9322 & 28.60 & 0.9373\tabularnewline
JORDER & 28.33 & 0.9088 & 29.24 & 0.9327 & 28.85 & 0.9340\tabularnewline
Ours model & 29.52 & 0.9278 & 29.33 & 0.9431 & 29.25 & 0.9482\tabularnewline
Ours model(RCAN) & 30.80 & 0.9339 & 29.62 & 0.9492 & 29.55 & 0.9532
     \end{tabular}%
 \end{table*}%
While there are methods directly learning a mapping function from RAW images to color images with neural networks, 
there is ground truth color images corresponding to RAW input since different ISP pipelines provide various RAW image processing algorithms. 
Hence, we need the color rainy image as complementary of RAW data limitation. The original rainy color image input can provide a reference to help recover the fidelity of color appearances. \\

Firstly, we pack the RAW image into 4 color channel and then adopt decomposition method\cite{DDN} to obtain RAW details. The RAW details then are fed into a deep CNN module for the purpose of joint negative rain streak feature learning and converting RAW data to RGB space. 
The module is followed by an upsampling layer with the pixel shuffle method\cite{PS}, as the demosaicking algorithm is approached via super resolution.  We also application decomposition method to extract color details.\\

 It is difficult to recover the relevant color corrections conducted within the ISP system, and thereby the networks trained with it could only be used for one specific camera. 
 Similar with \cite{RAWSR}, we adopt a shallow U-Net structure to predict color correction matrix with 12 channels output. This local color estimation contain a $ 3 \times 3 $ per pixel channel color matrix and $ 1 \times 3 $ offset color information. 
It is beneficial to combine the color details with the feature output from  JoRRRP module, this feature fusion make the color estimation module aware of the information from RAW details.  \\

\subsection{Backbone}
A straight forward backbone for the JoRRRP module is ResNet, but we need more backbones for ablation study of whether the RAW data information really help the rain removal task.
We deploy three more compact and efficient backbones: residual dense network (RDN)\cite{RDN},  residual channel attention network (RCAN)\cite{RCAN}, 
light weight information multi-distillation network (IMDN)\cite{IMDN}.\\

The RDN backbone exploit the hierarchical features by local feature fusion in RDB \cite{RDN}. After fully obtaining dense local features, the author then use global feature fusion to jointly and adaptively learn global hierarchical features. 
The RCAN backbone allows abundant low-frequency information to be bypassed through multiple skip connections by using residual in residual (RIR) structure. Furthermore, the network adaptively rescale features and considering inter dependencies among feature channels by channel attention mechanism.
This IMDN network construct the cascaded information multi-distillation blocks (IMDB), which contains distillation and selective fusion parts. In addition, the network propose contrast-aware channel attention mechanism according to the importance of candidate features.

\subsection{Implementation Details}

We obtain 2000 clean RAW images from MIT-Adobe 5K dataset \cite{fivek} to generate RAW and RGB training image pairs. The original dataset is composed of  high resolution RAW photos photographed from different kinds of light source.  We carefully remove  photos that is indoor, night or abstract patterns, which is not likely to be a rainy scene. We randomly selected 1800 images as training, 100 images as validation and 100 images as test. For each RAW image , we generate 4 RAW rain streak mask with different angles. The RAW data is resized to quarter of its original resolution to save computation. In addition, we separately choose  100 RAW images from HDR plus dataset and Fairchild dataset for testing the generalization ability. RAW images for testing is processed with different ISP algorithms containing different demosaicing methods, exposure scale and tone mapping methods.
\section{Experiments and Results}
\subsection{Results on Two RAW Synthetic Rain Dataset}
We first conduct experiments on our RAW synthetic rain data in order to clearly demonstrate
the importance of each component of our model. The result show that the performance is worsen without color branch, RAW input or feature fusion.

\subsection{Ablation study of backbones}

To evaluate the stability of our JoRRRP, we replace the ResN backbone with different backbones: RDN, RCAN and IMDN. The result shows that the performance of rain streak removal with RAW images is always better than RGB methods, with average 0.7 PSNR and 0.05 SSIM improvement.

\subsection{ISP settings}
 We also tested the performance of our solution with other RGB de-raining piplines and the results show that the model using RAW data have a better performance than models trained on RGB space images especially when the testing data is processed with different ISP pipelines.

\subsection{Comparison With The SOTA Methods}
We show comparisons about model size and performance in Figure 8. Although our RCAN is the deepest network, it has less parameter number than that of EDSR and RDN. Our RCAN and RCAN+ achieve higher performance, having a better tradeoff between model size and performance. It also indicates that deeper networks may be easier to achieve better performance than wider networks.

\section{Conclusions}
In this work, we propose a joint solution of rain streak removal and RAW processing. The results show that using RAW data achieves better accuracy and visual improvements against state-of-the-art methods on only RGB space images. In addition, we design 
We  present  a  technique  to  synthesize  rainy  image  on  RAW  data,  especially  pipeline  how  to  generate  rain  streak  on  RAW  data.  What is more, our method achieve better generalization performance than methods on color image when testing on different ISP pipelines.

\bibliography{egbib}

{\small
\bibliographystyle{ieee_fullname}
}

\end{document}